# The influence of nuclear models and Monte Carlo radiation transport codes on stray neutron dose estimations in proton therapy

M. De Saint-Hubert[1*], J. Farah[2], M. Klodowska[3], M. T. Romero-Expósito[4,5], K. Tyminska[6], V. Mares[7], P. Olko[8], L Stolarczyk[8,9] and S. Trinkl[10]

[1] Belgian Nuclear Research Centre (SCK CEN), Boeretang 200, BE-2400 Mol, Belgium.

[2] Institut de Radioprotection et de Sûreté Nucléaire (IRSN), Pôle Radioprotection de l'Homme, BP17, 92260 Fontenay-aux-Roses, France.

[3] Department of Medical Physics and Clinical Engineering, Addenbrooke's Hospital, Hills Road, Cambridge, CB2 0QQ, United Kingdom.

[4] Universitat Autònoma de Barcelona, Departament de Física, E-08193 Bellaterra, Spain.

[5] Instituto Tecnológico de Santo Domingo (INTEC), P.O. Box 342-9/249-2, Santo Domingo, República Dominicana.

[6] National Centre for Nuclear Research, A. Soltana 7, 05-400 Otwock-Swierk, Poland.

[7] Helmholtz Zentrum München, Institute of Radiation Medicine, Ingolstädter Landstraße 1, 85764 Neuherberg, Germany.

[8] Institute of Nuclear Physics PAN, Radzikowskiego 152, 31-342 Krakow, Poland.

[9] The Danish Centre for Particle Therapy, Aarhus University Hospital, Palle Juul-Jensens Boulevard 25, DK-8200 Aarhus, Denmark.

[10] Federal Office for Radiation Protection, Medical and Occupational Radiation Protection, Ingolstädter Landstraße 1, 85764 Neuherberg, Germany.

*corresponding author: mdsainth@sckcen.be

**Abstract**

**Purpose:** This study investigates the influence of several Monte Carlo radiation transport codes and nuclear models on the simulation of secondary neutron spectra and its impact on calculating and measuring the neutron doses in proton therapy.

**Materials and methods:** Three different multi-purpose Monte Carlo radiation transport codes (FLUKA, MCNPX, Geant4) were used together with different available nuclear models, to calculate secondary neutron energy spectra at various points inside a water tank phantom with PMMA walls using a 10 x 10 cm² rectangular, mono-energetic proton beam (110 MeV, 150 MeV, 180 MeV, 210 MeV). Using Kerma approximation secondary neutron doses were calculated applying fluence-to-dose equivalent conversion coefficients in water. Moreover, the impact of varying spectra for electrochemically etched CR39 detector calibration was analyzed for different codes and models.

**Results:** In distal positions beyond the Bragg peak, results show largest variations between the codes, which was up to 53% for the high energy neutron fluence at 16 cm from the Bragg peak of the 110 MeV proton beam. In lateral positions, the variation between the codes is smaller and for the total neutron fluence within 20%. Variation in the nuclear models in MCNPX was only visible for the proton beam energies of 180 and 210 MeV and modeling the high energy neutron fluence which reached up to 23% for 210 MeV at 11 cm lateral from the beam axis. Impact on neutron dose equivalent was limited for the different models used (<8%) while it was pronounced for the different codes (45% at 16 cm from the Bragg peak of the 110 MeV proton beam). CR39 calibration factors in lateral positions were on average varying 10% between codes and 5 % between nuclear models.

**Conclusions:** This study demonstrated a large impact on the neutron fluence spectra calculated by different codes while the impact of different models in MCNPX proved to be less prominent for the neutron modeling in proton therapy.

**Highlights:**

- Monte Carlo code and nuclear model impact neutron spectra, dose and CR39 calibration
- Codes have larger neutron spectra variation in distal versus lateral positions
- The influence of nuclear models on secondary neutron production is visible for 180 and 210 MeV proton beams
- Calculated neutron dose equivalent varied up to 45% between the codes
- CR39 calibration factors varied within 10% and 5% between codes and nuclear models

# Introduction

In recent years, major technological breakthroughs allowed for more compact and affordable proton therapy (PT) units, and a further increase the number and popularity of such facilities worldwide with more than 66 systems in operation in 2020 and 31 under construction. The physical behavior of protons result in a sharply localized peak of dose, known as the Bragg peak, allowing improved target dose conformation, with reduced entrance and negligible exit dose when compared to other radiotherapy techniques [1]. Nevertheless, one of the challenges of proton therapy is the production of secondary neutrons which are unavoidable due to nuclear interactions of high energy protons with beam line materials and with the patient's body [2]. As therapeutic proton beams have energies of hundreds of MeV and interact with materials of different tissue compositions and densities, the energy spectrum of these secondary neutrons range from thermal to the proton energy.

Currently, none of the available neutron counters and detectors is fully compatible with a clinical measurement of neutron spectra inside the patient or within an anthropomorphic phantom. Hence, many studies strongly rely on Monte Carlo (MC) particle transport calculations. The literature extensively reports the use of multiple purpose MC codes such as FLUKA [3-5], MCNPX [6], GEANT4 [7] and PHITS [8] for several applications in proton therapy. First of all, MC simulations can be used for out-of-field dosimetry as it allows to compute neutron doses [9-11], which are not considered by the current treatment planning systems used in PT. Furthermore, the shielding of proton therapy facilities is often based on results from MC simulations which allow the computation of neutron ambient dose equivalent [12-14] and the spectral neutron fluence inside and outside the treatment room [15-19].

Only few studies have compared MC simulations to neutron doses, ambient dose equivalents and Bonner sphere spectrometry measured in proton therapy [16, 19-24]. Such studies reported large discrepancies between experimental results and MC simulations by up to factors 2-3, presumably due to large measurement uncertainties as well as limitations of nuclear reaction models and cross sections used in the MC codes. In general, MC codes allow accurate calculations for neutrons below 20 MeV thanks to existing and well evaluated data libraries, such as ENDF/B [25, 26], which provide reliable neutron cross section data. Above 20 MeV, cross-section data are scarce or non-existing for several materials and MC codes. Up to 150 MeV neutrons, MCNP has the ability to utilize data libraries that have recently been released by LANL Group T-2 [27]. Nevertheless, some codes do not use these cross-sections for higher energies and need to rely on nuclear models that describe the interaction of protons and neutrons with target nuclei. Several of these models are available such as Intranuclear Cascade (INC) models (e.g. Bertini, Binary INC model, ISABEL model), pre-equilibrium models as well as evaporation models (e.g. Dresner and Abla). In general, it is difficult to define which of the models are more suitable for simulations in a particular application and for specific elements. Benchmarking studies have been conducted for heavier elements such as copper and iron [28, 29] but not yet for light elements constituting biological tissues such as hydrogen, oxygen, nitrogen, carbon. Moreover, the influence of MC codes and selection of nuclear models have been tested for Bonner sphere spectrometry (BSS) measurements, since the determination of the neutron spectrum by unfolding requires the input of an initial guess spectrum. It was shown that secondary neutron doses from cosmic irradiation as measured with BSS have an uncertainty of 10 % related to the different nuclear models and transport codes (GEANT4 and MCNP) [30]. More recently, an even more extended intercomparison of codes (MCNP, MCNPX, FLUKA, PHITS, MARS, or GEANT4) showed an uncertainty of unfolded neutron fluences above 20 MeV of about 20 % [31]. Not only BSS and rem counters require MC simulations to assess pre-requisite information for their calibration, also passive detector systems may require MC simulations for energy response correction and/or appropriate calibration. For example, the calibration of electrochemically-etched track detectors (CR39), i.e. the conversion of track density into dose, relies on a fluence factor, which is often estimated through MC simulations for a predefined standard neutron source [32, 33]. The accuracy and consistency of MC simulations may hence affect experimental measurements by expanding their associated uncertainties and adding a major component which is currently not quantified.

The European Radiation Dosimetry Group working group 9 (EURADOS WG9) research focusses on the assessment of neutron ambient dose in the proton treatment room and in the facility [34-36] as well as in the patient, more specifically assessing the undesired out-of-field doses during proton therapy [37-39]. Several types of ambient radiation monitors as well as numerous passive detector types have been studied and compared for stray radiation using water and anthropomorphic phantoms. Also comparison of experimental data to MC calculations is often performed, involving the use of many different MC codes and models. Nevertheless, an intercomparison of the different available Monte Carlo codes is missing and needed to assess their performance, identify limitations related to the use of MC codes as well as the influence the selection of a certain MC code can have on calibration of passive detector systems used during experimental data. This study focused on comparing three widely used MC codes, FLUKA, MCNPX and GEANT4, in the prediction of secondary neutrons following nuclear reactions of a typical proton therapy beam with light elements. The work first involved modeling of a large experimental campaign performed by WG9 [36, 37]. First, MC codes were compared in reproducing the range of therapeutic pencil proton beams targeting a 30 x 60 x 30 cm³ water tank phantom. Next, simulations of neutron spectra inside the water tank phantom were performed at different depths and lateral positions with respect to the Bragg peak and the different fluences were compared. Finally, the variability of neutron spectra among the codes and the influence on the calculated calibration factor was assessed for electrochemically etched CR-39 detectors used in a previously conducted experimental campaign [36, 37]. The spectra are needed to determine the calibration factor because in electrochemically etched CR-39 detectors the size of a track does not depend on neutron energy [33].

## Materials and Methods

### Water phantom and beam parameters

A 30 x 60 x 30 cm³ water phantom with polymethyl methacrylate (PMMA) wall thickness of 15 mm and a beam entrance wall with thickness of 4 mm (area 12 x 12 cm$^2$) was modelled as shown in figure 1. This water phantom was developed by Bordy, et al. [40] and used during previously conducted experimental studies within EURADOS WG9 [36, 37]. To investigate the influence of nuclear models on MC particle transport calculations, a simple beam model was implemented with four different proton energies of 110 MeV, 150 MeV, 180 MeV and 210 MeV. A 10 x 10 cm² rectangular parallel beam of mono-energetic protons was modelled entering the water phantom at the beam entrance window. Outside the water phantom the beam was travelling through 50 cm of air.

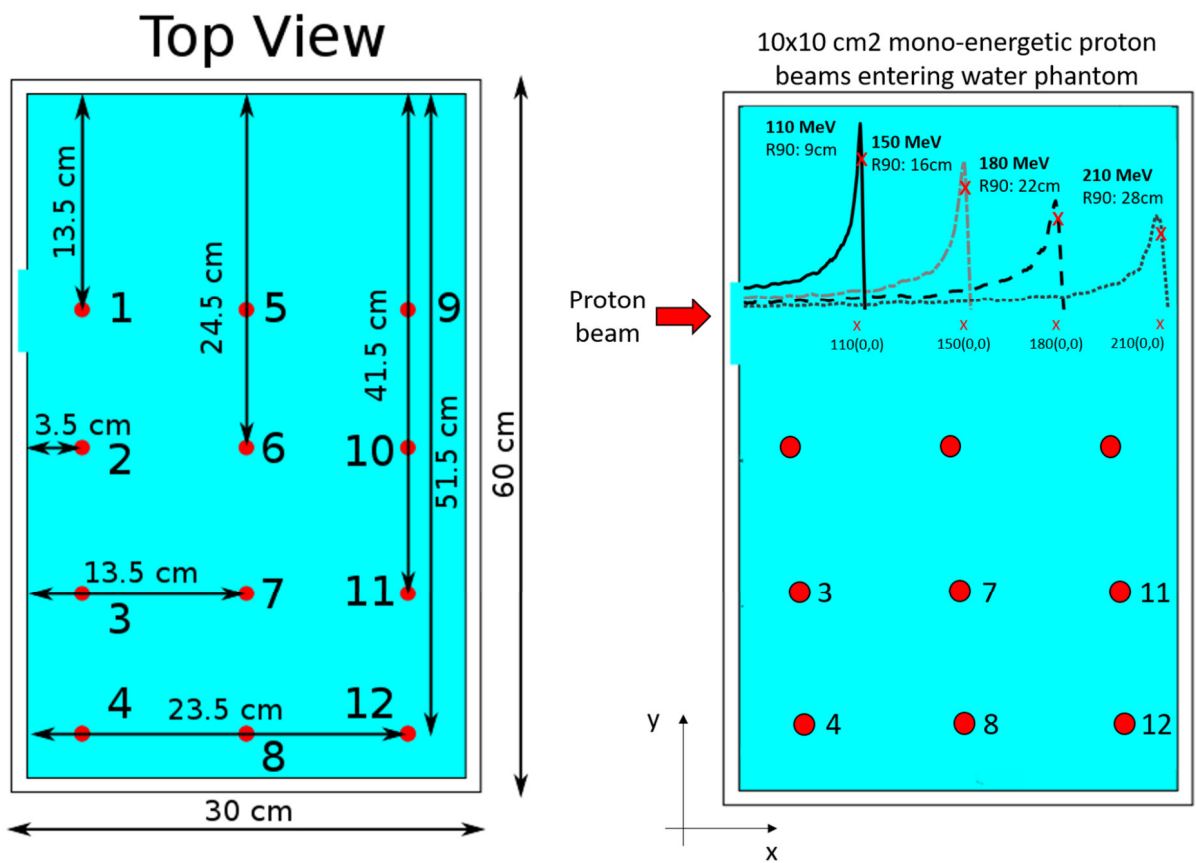

**Figure 1.** The left-hand side figure shows a schematic representation of the 30 x 60 x 30 cm³ water phantom with PMMA walls of 15 mm consisting of a beam entrance window with a thickness of only 4 mm (area 12 x 12 cm²) and 12 different positions (1 cm diameter spheres) used for the calculations performed with the different MC simulation codes/models .On the right-hand side, the entrance of the parallel rectangular 10x10 cm$^2$ proton beams indicated with an arrow including Bragg curves for the respective proton energies (MCNPX) demonstrating their different ranges (R90 values) and positions with respect to the 12 positions in the water phantom.

First, the depth dose distribution for all the four mono-energetic beams was scored using voxelization of the water phantom, with cubic voxel sizes of 1 mm³. For the definition of range, the 90% dose in the distal falloff (R90) values were calculated and we report in this manuscript on the interpolated R90 values for each energies. The distance refers to the distance inside the water phantom (outer wall is set to position zero), thus including a 4 mm PMMA wall followed by water (see figure 1).

Furthermore 12 different positions were defined inside the water phantom for tallying of neutron spectra, neutron dose equivalent and CR39 calibration factor in spherical volumes of 1 cm diameter (see figure 1). The relative positions to the proton beams' isocenters are specified in table 1.

**Table 1**. Overview of the relative position within the radiation field of positions 1, 5, 9, 2, 6 and 10 for the different mono-energetic proton beams. Bold font indicates that we consider out-of-field, i.e. not within the proton beam's Bragg curve. The numbers in brackets are the distances (in cm) to the Bragg peak (R90) isocenters (0,0) in x and y coordinates as indicated in figure 1.

| Positions | Mono-energetic proton beams | | | |
|---|---|---|---|---|
| | 110 MeV | 150 MeV | 180 MeV | 210 MeV |
| 1 | Front Bragg peak (-4,0) | Front Bragg peak (-10,0) | Front Bragg peak (-17,0) | Front Bragg peak (-23,0) |
| 5 | **Distal Bragg peak** (6,0) | Bragg peak (0,0) | Plateau Bragg peak (-7,0) | Plateau Bragg peak (-13,0) |
| 9 | **Distal Bragg peak** (16,0) | **Distal Bragg peak** (10,0) | **Distal Bragg peak** (3,0) | Plateau Bragg peak (-3,0) |
| 2 | **Lateral Bragg peak** (-4,11) | **Lateral Bragg peak** (-10,11) | **Lateral Bragg peak** (-17,11) | **Lateral Bragg peak** (-23,11) |
| 6 | **Lateral Bragg peak** (6,11) | **Lateral Bragg peak** (0,11) | **Lateral Bragg peak** (-7,11) | **Lateral Bragg peak** (-13,11) |
| 10 | **Lateral Bragg peak** (16,11) | **Lateral Bragg peak** al (10,11) | **Lateral Bragg peak** (3,11) | **Lateral Bragg peak** (-3,11) |

Identical material composition and densities for PMMA, air and water (shown in table 2) were used in all codes. Densities were considered for room temperature.

Table 2. Material densities and compositions used for the MC calculations

| Material | Density (g/cm³) | Isotope | Mass fraction |
|---|---|---|---|
| Air (dry) | 0.001205 | $^{14}$N | 0.75527 |
| | | $^{40}$Ar | 0.01282 |
| | | $^{16}$O | 0.23178 |
| | | $^{12}$C | 0.00012 |
| Water | 0.998 | $^{16}$O | 0.66666 |
| | | $^{1}$H | 0.33333 |
| PMMA | 1.18 | $^{12}$C | 0.59985 |
| | | $^{16}$O | 0.31962 |
| | | $^{1}$H | 0.08054 |

**Monte Carlo codes and nuclear models**

Each participant, using a specific Monte Carlo code (1 for GEANT4, 4 for MCNPX and 1 for FLUKA), created dedicated input files for modeling the neutron energy spectra in the different positions in the water phantom. The neutron fluence data calculated by the different codes were analyzed and compared for four neutron energy regions including thermal (E < 0.4 eV), epithermal (0.4 eV < E ≤ 100 keV), fast (100 keV < E ≤ 19.6 MeV) and high energy (E > 19.6 MeV) neutrons. The variation between the codes was assed by calculating the coefficient of variation between the codes (i.e. ratio of standard deviation to the mean). The number of simulated particles from the source was sufficient to assure statistical uncertainties, coverage factor k=1, in the bins should be below 5% while for the fluence, dose calculations and calibration factors statistical uncertainties had to remain within 3%.

GEANT4

The binary intra-nuclear cascade model (BIC) [41] was used by setting the standard physics lists QGSP_BIC_HP. Furthermore, the physics list was modified with the electromagnetic physics option 3 and extended for the treatment of thermal neutrons with the G4NeutronHPThermalScattering physics. In order to use the thermal scattering physics for hydrogen in water, it was necessary to use an element called "TS_H_of_Water" defined in G4NeutronHPThermalScatteringNames.cc, that allows thermal scattering cross sections to be activated for water. For all simulations, GEANT4 version 10.1.2 was used and neutron fluence was scored per energy bin using path length estimation (G4PSCellFlux primitive scorer) on spherical volumes. We have used 10 log-equidistant bins per decade in the whole energy range from 8.91E-10 MeV to 2.20E+02 MeV.

MCNPX

The Monte Carlo N-Particle eXtended (MCNPX) transport code version 2.7.0 [42] was used in this exercise by 4 different institutes allowing a comparison between the MC output for the same code as implemented by the 4 different groups. The Los Alamos LA150H and LA150N cross section data libraries were used respectively for protons and neutrons [43, 44]. These cross section data libraries are evaluated for about 40 target isotopes and for incident proton energies ranging from 1 MeV to 150 MeV and neutrons from 20 MeV to 150 MeV [45]. Below 20 MeV neutrons and 1 MeV protons endf/b-vii.0 was used. For carbon and argon, models were used for protons, which is expected to have a small impact due to the limited presence of these elements in the water phantom."Below 20 MeV neutrons and 1 MeV protons endf/b-vii.0 was used. Due to the fact that reaction cross section libraries are not available above 150 MeV, the Bertini (Bert) intra-nuclear cascade (INC) model [46] and the Dresner (Dres) evaporation-fission model [47] was used (default setting in MCNPX). In addition to the default Bertini-Dresner model, different combinations of Bertini and Isabel (Isa) [48] (for INC modeling) together with Dresner and Abla models (for evaporation phase) were considered. Each participant simulated neutron spectra, using 10 log-equidistant bins per decade in the whole energy range from 8.91E-10 MeV to 2.20E+02 MeV. Applying the F4-type tally, for Bert-Dres, Bert-Abla, Isa-Abla and Isa-Dres as well as considering the Cascade-Exciton Model (CEM version 03), combining essential features of the excition and INC models [49, 50]. For a proper evaluation of thermal neutrons, room temperature cross section tables S(a,b) in water (lwtr.10t) were included based on ENDF/B-VII.0 [10].

FLUKA

Physics in FLUKA is unique and unchangeable regardless of chosen settings influencing only code efficiency and calculation precision [51]. For this study, the 2011.2. FLUKA version was used and the HADROTHErapy default settings were applied. This implicates a particle transport threshold at 100 keV except for neutrons simulated down to thermal energies [52]. The PEANUT package is incorporated for hadron inelastic nuclear interactions [53] and the modified RQMD (Relativistic Quantum Molecular Dynamic) model [54] is employed for nucleus-nucleus interactions between 0.125 and 5 GeV, while below 125 MeV the Boltzmann Master Equation (BME) model is used [55]. For the neutron dose calculation, the USRBIN card was switched on for each point of interest (a sphere with 1 cm diameter), whereas for neutron spectra a USRTRACK card was used within the water phantom multichannel. The binning of low-energy neutron groups up till 20 MeV is pre-defined using a multi-group algorithm based on built-in cross section data library binned into 260 energy groups (not log-equidistant), 31 of which are in the thermal energy region [5]. For energies above 20 MeV, 10 log-equidistant bins per decade were used up to 2.20E+02 MeV in the same manner as for the other 2 codes used in this study.

**Neutron dose equivalent calculation from neutron spectra**

The impact of modeling the neutron spectra on the simulated neutron dose equivalent for different MC codes and nuclear models was evaluated by using the method explained by Romero-Expósito et al. [32]. Assuming the validity of the kerma approximation, the absorbed dose can be approximated by kerma which, in turn, may be evaluated from neutron fluence through the kerma factors $k(E)$ for ICRU tissue found in the work of *Siebert and Schuhmacher* [56] for neutrons up to 20 MeV and in the work of *Chadwick et al.* up to 150 MeV [44]. Applying the neutron quality factor as a function of energy ($Q(E)$), the neutron dose equivalent can be derived using the following equation:

$$H = \int_E Q(E) \cdot k(E) \cdot \frac{d\phi_i(E)}{dE} \cdot dE \qquad (1)$$

where $\frac{d\phi_i(E)}{dE}$ is the energy spectrum of the absolute neutron fluence.

**Neutron dose equivalent measurements from CR 39 passive detectors**

The basis of dose equivalent evaluation relies on the same equation 1 with a small modification:

$$H = \Phi \int_E Q(E) \cdot k(E) \cdot \frac{d\varphi(E)}{dE} \cdot dE \quad H = \Phi \int_E Q(E) \cdot k(E) \cdot \frac{d\varphi_i(E)}{dE} \cdot dE \qquad (2)$$

$H = \Phi \int_E Q(E) \cdot k(E) \cdot \frac{d\varphi_i(E)}{dE} \cdot dE$ where $\Phi$ is the total neutron fluence, and is obtained by the CR39 passive detector, and $\frac{d\varphi(E)}{dE}$, the energy spectrum of the unit neutron fluence.

As explained in Romero-Expósito et al. [32], total fluence can be evaluated from a CR39 reading (*N*) taking into account an average response factor which considers the fractions of the different components of the neutron spectrum:

$$\Phi = \frac{N}{\overline{R_\Phi}} = \frac{N}{p_{epi+th} \cdot R_{epi+th} + p_{fast} \cdot R_{fast} + p_{high} \cdot R_{high}} \quad (3)$$

$p_{epi+th}$, $p_{fast}$, and $p_{high}$ denotes for the thermal and epithermal fraction (1 meV < E ≤ 100 keV), fast fraction (100 keV < E ≤ 19.6 MeV), and high energy fraction (E > 19.6 MeV),, respectively, and $R_{epi+th}$, $R_{fast}$ and $R_{high}$ are the corresponding fluence responses.

Combination of equations 2 and 3 allows to derive the expression used for estimation of the calibration coefficient:

$$\left(H/N\right) = \frac{\int_E Q(E) \cdot k(E) \cdot \frac{d\varphi_i(E)}{dE} \cdot dE}{p_{epi+th} \cdot R_{epi+th} + p_{fast} \cdot R_{fast} + p_{high} \cdot R_{high}} \quad (4)$$

# Results

## Benchmarking proton beam

As a first step, a general validation on proton beam ranges was carried out, for all the four mono-energetic beams. In Table 3 the interpolated R90 values from the depth dose distribution are shown measured from the outer wall of the phantom. A good agreement between the codes was found with the largest differences in R90 value of 1.1 mm for the 180 MeV proton beam. This was considered sufficient for the purpose of this study.

**Table 3.** R90 values calculated with 3 different Monte Carlo codes (default settings) using 10 x 10 cm² parallel beam of mono-energetic protons.

| R90 (cm) | 110 MeV | 150 MeV | 180 MeV | 210 MeV |
|---|---|---|---|---|
| **GEANT4** | 8.98 | 15.62 | 21.58 | 28.06 |
| **MCNPX** | 9.01 | 15.62 | 21.51 | 28.07 |
| **FLUKA** | 8.98 | 15.62 | 21.47 | 28.02 |

## Neutron spectra of different MC codes inside the phantom

First we compared the different neutron spectra for the different codes (MCNPX, GEANT4 and FLUKA). For MCNPX we used the default nuclear model (Bert-Dres) and we reported on the data of 1 participant.

*110 MeV proton beam*

Figure 2 shows the results on simulated secondary neutron energy spectra inside the water phantom for 110 MeV protons at distal positions 5 and 9 and lateral positions 6 and 10. All spectra in Figure 2 have similar shape consisting of two main components (peaks) at the thermal (E < 0.4 eV) and fast and high energy range (E > 0.1 MeV). When comparing these spectra it can be seen that the heights of peaks are different depending on the position with respect to beam line and Bragg peak. Moreover the influence of different nuclear models and MC codes on secondary neutron spectra is most prominent in the energy range above ~5 MeV.

Tables 4 summarizes the fluence data calculated by the different codes (default models) for 110 MeV at distal positions 5 and 9 and lateral positions 2, 6 and 10 in four neutron energy regions of thermal (E < 0.4 eV), epithermal (0.4 eV < E ≤ 100 keV), fast (100 keV < E ≤ 19.6 MeV) and high energy (E > 19.6 MeV). Variation between the codes was calculated by estimating the relative standard deviation to the mean of the different codes. Results show that the thermal neutron fluences vary within 11% while fluences for the fast and high energy neutrons vary within 30% and 53%, respectively. In the forward scattering directions, i.e. at position 5 and 9, variations in high energy neutron fluence were found up to 46 % and 53 %, respectively, and variations on the total neutron fluence of 16% and 29%, respectively. In lateral positions 2, 6 and 10 the variation on the high energy neutron fluence was found to be lower than in forward directions, up to 12%, 22% and 36%, respectively, while variation on the total neutron fluence was found to be 14%, 6% and 19%, respectively.

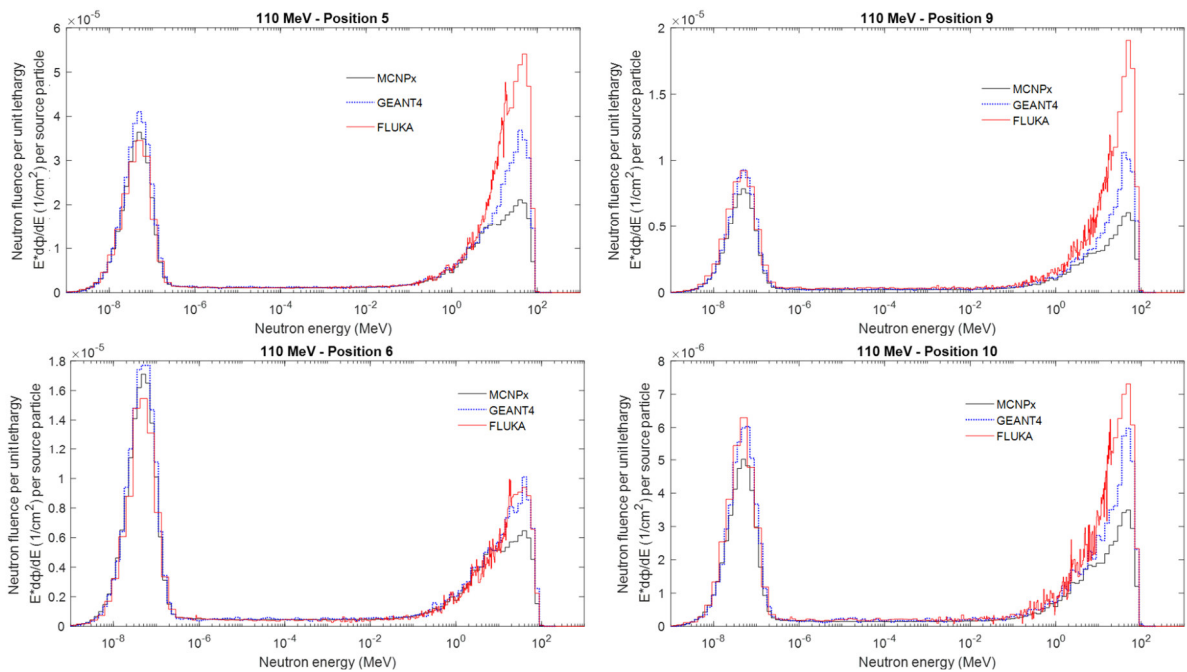

**Figure 2.** Neutron spectra simulated with different MC codes for a 110 MeV proton beam at distal positions 5 and 9 and lateral positions 2 and 10 inside the water phantom. (FLUKA binning was smaller than GEANT4 and MCNPX up to 20 MeV)

For comparison the difference to the MCNPX results was calculated for the GEANT4 and FLUKA codes (see table 4). This was done because MCNPX uses evaluated nuclear cross-sections up to 150 MeV provided in LA150H and LA150N for several materials. Moreover MCNPX has been used by 4 different participants, demonstrating very good agreement, and also particularly since we looked more explicitly into influence of nuclear models on secondary neutron production for MCNPX models (see section MCNPX intercomparison).

The results from GEANT4 simulations showed a higher thermal neutron fluence compared to MCNPX for all positions except for position 2. This difference was largest for position 10 (17%). FLUKA simulations-based estimations of the thermal neutron fluence were within 10% for positions 5 and 6 while a lower fluence was observed for position 2 (-17%) and a higher fluence for positons 9 (25%) and 10 (21%). Moreover, the data show that the FLUKA code gave higher fluence for the high energy neutrons in forward direction compared to the results obtained with MCNPX. This was 162% in position 5 and up to 200% for position 9. The simulations with GEANT4, on the other hand, also produced higher fluence for the high energy neutrons, namely up to 72% in position 5.

**Table 4.** Secondary neutron fluence (1/cm$^2$) per source particle in distal position 5 and 9 and lateral positions 2, 6 and 10 for 110 MeV proton beam distinguishing between Thermal (E ≤ 0.4 eV), Epithermal (0.4 eV < E ≤ 100 keV), Fast (100 keV < E ≤ 19.6 MeV) and High energy neutrons (E > 19.6 MeV).

|  |  | MCNPX | GEANT4 |  | FLUKA |  | Variation between codes (%) |
|---|---|---|---|---|---|---|---|
|  |  | Bert-Dres | BIC | Difference to MCNPX | HADROTHE | Difference to MCNPX |  |
| Position 5 | Thermal | 6.96E-05 | 7.64E-05 | 10% | 6.84E-05 | -2% | 6% |
|  | Epith | 1.46E-05 | 1.56E-05 | 6% | 1.51E-05 | 3% | 3% |
|  | Fast | 4.42E-05 | 5.15E-05 | 17% | 6.48E-05 | 47% | 20% |
|  | High | 2.39E-05 | 4.11E-05 | 72% | 6.27E-05 | 162% | 46% |
|  | TOTAL | 1.52E-04 | 1.85E-04 | 21% | 2.11E-04 | 38% | 16% |
| Position 9 | Thermal | 1.50E-05 | 1.71E-05 | 13% | 1.88E-05 | 25% | 11% |
|  | Epith | 3.07E-06 | 3.43E-06 | 12% | 4.28E-06 | 39% | 17% |
|  | Fast | 9.16E-06 | 1.14E-05 | 24% | 1.65E-05 | 80% | 30% |
|  | High | 6.74E-06 | 1.15E-05 | 70% | 2.02E-05 | 200% | 53% |
|  | TOTAL | 3.40E-05 | 4.34E-05 | 28% | 5.98E-05 | 76% | 29% |
| Position 2 | Thermal | 4.39E-05 | 4.30E-05 | -2% | 3.64E-05 | -17% | 10% |
|  | Epith | 1.03E-05 | 9.60E-06 | -7% | 7.29E-06 | -29% | 17% |
|  | Fast | 2.30E-05 | 2.10E-05 | -9% | 1.49E-05 | -35% | 22% |
|  | High | 4.09E-06 | 4.82E-06 | 18% | 3.82E-06 | -7% | 12% |
|  | TOTAL | 8.13E-05 | 7.84E-05 | -4% | 6.24E-05 | -23% | 14% |
| Position 6 | Thermal | 3.26E-05 | 3.41E-05 | 5% | 2.99E-05 | -8% | 7% |
|  | Epith | 5.98E-06 | 6.33E-06 | 6% | 5.52E-06 | -8% | 7% |
|  | Fast | 1.58E-05 | 1.77E-05 | 12% | 1.60E-05 | 2% | 6% |
|  | High | 7.02E-06 | 1.01E-05 | 44% | 1.09E-05 | 55% | 22% |
|          | TOTAL   | 6.14E-05 | 6.82E-05 | 11% | 6.24E-05 | 2%   | 6%  |
| Position 10 | Thermal | 9.59E-06 | 1.12E-05 | 17% | 1.16E-05 | 21%  | 10% |
|          | Epith   | 1.95E-06 | 2.31E-06 | 19% | 2.54E-06 | 31%  | 13% |
|          | Fast    | 5.70E-06 | 7.20E-06 | 26% | 8.70E-06 | 53%  | 21% |
|          | High    | 3.84E-06 | 6.21E-06 | 62% | 8.30E-06 | 116% | 36% |
|          | TOTAL   | 2.11E-05 | 2.69E-05 | 28% | 3.12E-05 | 48%  | 19% |

*210 MeV proton beam*

Figure 3 shows the neutron spectra in lateral positions 2, 6 and 10 for a 210 MeV proton beam, as calculated by the different codes (default models). Two main peaks can be observed; higher thermal peak and lower fast-high peak above 100 keV. Interestingly, the influence of different MC codes and models is more pronounced in thermal peaks, while spectra of secondary neutrons above 100 keV are rather similar. Table 5 shows a variation between codes on the average data of the codes of around 20%, for all simulation positions. The data suggest that the high energy neutron fluence obtained from the simulations showed less variation between the codes (14%, 9% and 4% for positions 2, 6 and 10, respectively) than the lower energy neutron fluence. Moreover, simulation results at position 10 seem to have a better agreement. Looking into GEANT4 and FLUKA all these positions showed lower dose estimations when compared to MCNPX except for the high energy neutrons in position 10.

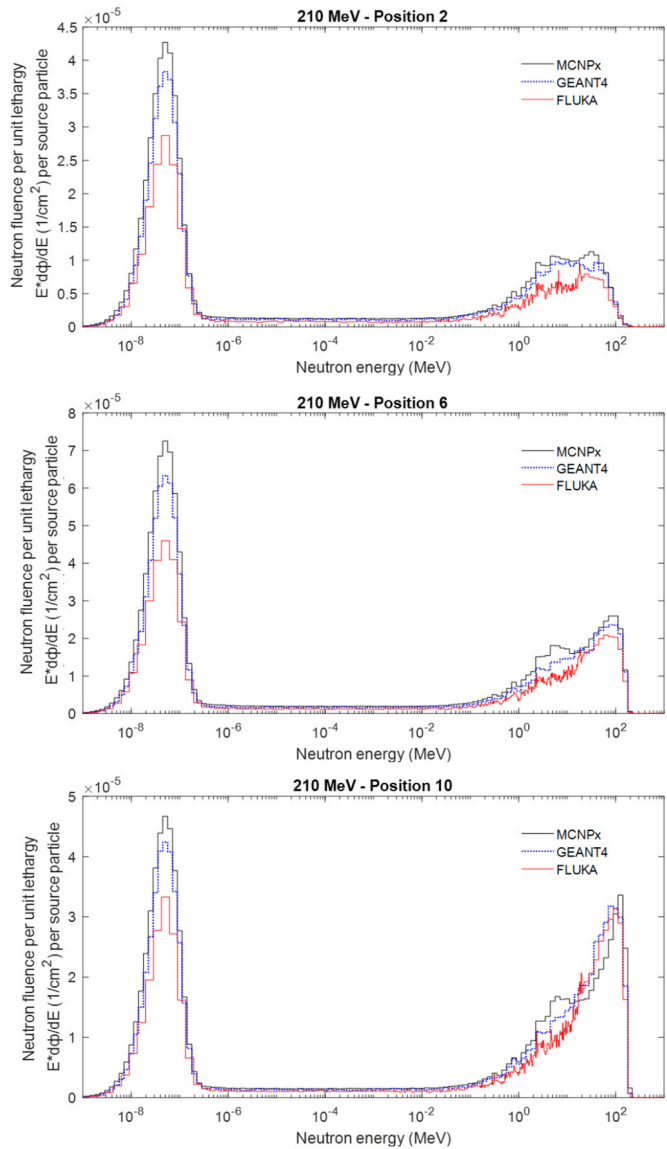

**Figure 3.** Neutron spectra simulated with different MC codes for 210 MeV proton beams at lateral positions 2, 6 and 10 inside the water phantom. (FLUKA binning was smaller than GEANT4 and MCNPX up to 20 MeV)

*Comparison of 110 MeV and 210 MeV proton beams*

To compare data for different proton beams at a similar location, position 2 for 110 MeV (table 4) with positon 10 for 210 MeV were considered (see table 5). As table 2 shows, the x, y positions relative to the isocenter are (-4 cm, 11 cm) in position 2 for the 110 MeV proton beam and (-3 cm, 11 cm) in position 10 for the 210 MeV proton beam. In both positions a similar spectral shape is observed (figure 2 bottom left plot and figure 3 top plot) but the maximum neutron energy is different due to the difference in proton energy.

When comparing variations in the total fluence, these were found to be very similar: 14% for 110 MeV in positon 2 and 12% for 210 MeV in position 10.

In both energies and respective positions, lower fluence values were obtained from GEANT4 simulations than from MCNPX simulations for low-energy neutrons while higher fluence data are observed for MCNPX in the high-energy neutron fluence spectrum compared to GEANT4. For 110 MeV at position 2 the difference is up to 18% higher for MCNPX compared to GEANT4. FLUKA simulation results gave a 22 % and 23 % lower total neutron fluence, compared to MCNPX results for 110 MeV and 210 MeV, respectively, at positions 2 and 10.

**Table 5.** Secondary neutron fluence (1/cm$^2$) per source particle in lateral positions 2, 6 and 10 for 210 MeV proton beam distinguishing between Thermal (E ≤ 0.4 eV), Epithermal (0.4 eV < E ≤ 100 keV), Fast (100 keV < E ≤ 19.6 MeV) and High energy neutrons (E > 19.6 MeV). Variation between the codes was calculated as…

|  |  | **MCNPX** | **GEANT4** | | **FLUKA** | | Variation between codes (%) |
|---|---|---|---|---|---|---|---|
|  |  | Bert-Dres | BIC | Difference to MCNPX | HADROTHE | Difference to MCNPX |  |
| Position 2 | Thermal | 8.18E-05 | 7.19E-05 | -12% | 5.53E-05 | -32% | 19% |
| | Epith | 1.67E-05 | 1.46E-05 | -13% | 1.02E-05 | -39% | 24% |
| | Fast | 3.54E-05 | 3.16E-05 | -11% | 2.12E-05 | -40% | 25% |
| | High | 1.42E-05 | 1.27E-05 | -10% | 1.08E-05 | -24% | 14% |
| | TOTAL | 1.48E-04 | 1.31E-04 | -12% | 9.76E-05 | -34% | 20% |
| Position 6 | Thermal | 1.38E-04 | 1.19E-04 | -14% | 9.11E-05 | -34% | 20% |
| | Epith | 2.56E-05 | 2.15E-05 | -16% | 1.60E-05 | -37% | 23% |
| | Fast | 5.74E-05 | 4.81E-05 | -16% | 3.52E-05 | -39% | 24% |
| | High | 4.46E-05 | 4.13E-05 | -7% | 3.76E-05 | -16% | 9% |
| | TOTAL | 2.66E-04 | 2.30E-04 | -14% | 1.80E-04 | -32% | 19% |
| Position 10 | Thermal | 8.92E-05 | 7.95E-05 | -11% | 6.22E-05 | -30% | 18% |
| | Epith | 1.97E-05 | 1.71E-05 | -13% | 1.36E-05 | -31% | 18% |
| | Fast | 5.08E-05 | 4.56E-05 | -10% | 3.58E-05 | -30% | 17% |
| | High | 4.96E-05 | 5.38E-05 | 8% | 5.26E-05 | 6% | 4% |
| | TOTAL | 2.09E-04 | 1.96E-04 | -6% | 1.64E-04 | -22% | 12% |

**MCNPX intercomparison and influence of nuclear models on secondary neutron production**

In this study, four MCNPX implementations in four different institutes were used. Interestingly all participants used the same version of MCNPX 2.7 and the differences between the fluence spectra in the different energy windows were within 5% for the same nuclear model settings. In the following, results from 3 distinct models, namely Bert-Dres, CEM and Isa-Able, presented largest deviations when analyzing their respective neutron spectra.

For 110 MeV and 150 MeV protons the simulated secondary neutron fluences varied only within 2%. These results were expected because in energy region below 150 MeV the high-energy cross section data libraries were used instead of high-energy models.

**Table 6.** Secondary neutron fluence (1/cm$^2$) per source particle for different MCNPX nuclear models in positions 2, 6 and 10 for 180 and 210 MeV proton beams distinguishing between Thermal (E ≤ 0.4 eV), Epithermal (0.4 eV < E ≤ 100 keV), Fast (100 keV < E ≤ 19.6 MeV) and High energy neutrons (E > 19.6 MeV).

|  |  | 180MeV | | | Variation between models (%) | 210MeV | | | Variation between models (%) |
|---|---|---|---|---|---|---|---|---|---|
|  |  | Bert-Dres | CEM | Isa-Abla |  | Bert-Dres | CEM | Isa-Abla |  |
| Position 2 | Thermal | 8.49E-05 | 8.84E-05 | 8.69E-05 | 2% | 8.15E-05 | 9.27E-05 | 8.56E-05 | 7% |
|  | Epith | 1.77E-05 | 1.85E-05 | 1.82E-05 | 2% | 1.66E-05 | 1.90E-05 | 1.75E-05 | 7% |
|  | Fast | 3.59E-05 | 3.72E-05 | 3.69E-05 | 2% | 3.34E-05 | 3.90E-05 | 3.53E-05 | 8% |
|  | High | 1.39E-05 | 1.06E-05 | 1.65E-05 | 22% | 1.66E-05 | 1.25E-05 | 2.00E-05 | 23% |
|  | TOTAL | 1.52E-04 | 1.55E-04 | 1.59E-04 | 2% | 1.48E-04 | 1.63E-04 | 1.58E-04 | 5% |
| Position 6 | Thermal | 1.22E-04 | 1.18E-04 | 1.22E-04 | 2% | 1.38E-04 | 1.36E-04 | 1.33E-04 | 2% |
|  | Epith | 2.29E-05 | 2.24E-05 | 2.30E-05 | 1% | 2.55E-05 | 2.49E-05 | 2.40E-05 | 3% |
|  | Fast | 5.04E-05 | 4.71E-05 | 5.07E-05 | 4% | 5.41E-05 | 5.34E-05 | 5.20E-05 | 2% |
|  | High | 3.66E-05 | 2.96E-05 | 3.64E-05 | 12% | 4.85E-05 | 3.39E-05 | 4.50E-05 | 18% |
|  | TOTAL | 2.32E-04 | 2.17E-04 | 2.33E-04 | 4% | 2.66E-04 | 2.49E-04 | 2.54E-04 | 3% |
| Position 10 | Thermal | 5.58E-05 | 5.58E-05 | 5.57E-05 | 0% | 8.88E-05 | 8.84E-05 | 7.99E-05 | 6% |
|  | Epith | 1.17E-05 | 1.16E-05 | 1.16E-05 | 0% | 1.95E-05 | 1.94E-05 | 1.87E-05 | 2% |
|  | Fast | 3.00E-05 | 2.82E-05 | 2.98E-05 | 3% | 4.76E-05 | 4.49E-05 | 4.76E-05 | 3% |
|  | High | 3.46E-05 | 2.97E-05 | 3.30E-05 | 8% | 5.34E-05 | 4.31E-05 | 5.16E-05 | 11% |
|  | TOTAL | 1.32E-04 | 1.25E-04 | 1.30E-04 | 3% | 2.09E-04 | 1.96E-04 | 1.98E-04 | 4% |

.

Table 6 and figure 4 show neutron fluences obtained with MCNPX simulations considering 3 different models Bert-Dres, CEM and Isa-Abla for primary proton energies of 180 MeV and 210 MeV at 3 different positions 2, 6 and 10. The effect of the models on the fluence of thermal, epithermal and fast neutrons are small with variations within 8%. However, a clear difference between the three considered nuclear models is observed for the high energy neutron fluence. At position 2, the high energy neutron fluence variation between models was 22% and 23% for 180 MeV and 210 MeV protons, respectively. Position 10 shows lower variations in high energy neutron fluence for different nuclear models compared to position 2, with variation between the models of 8% and 11% for 180 MeV and 210 MeV, respectively.

In general the variation between the models is higher for 210 MeV proton beams compared to 180 MeV protons, except for position 2, where a similar variation between the models is observed 22% and 23% for 180 MeV and 210 MeV, respectively, is observed (cf. figure 3 and table 6).

Comparing the three models, Bert-Isa-Dres produces similar results as Isa-Abla, while CEM gives a lower fluence for high energy neutrons.

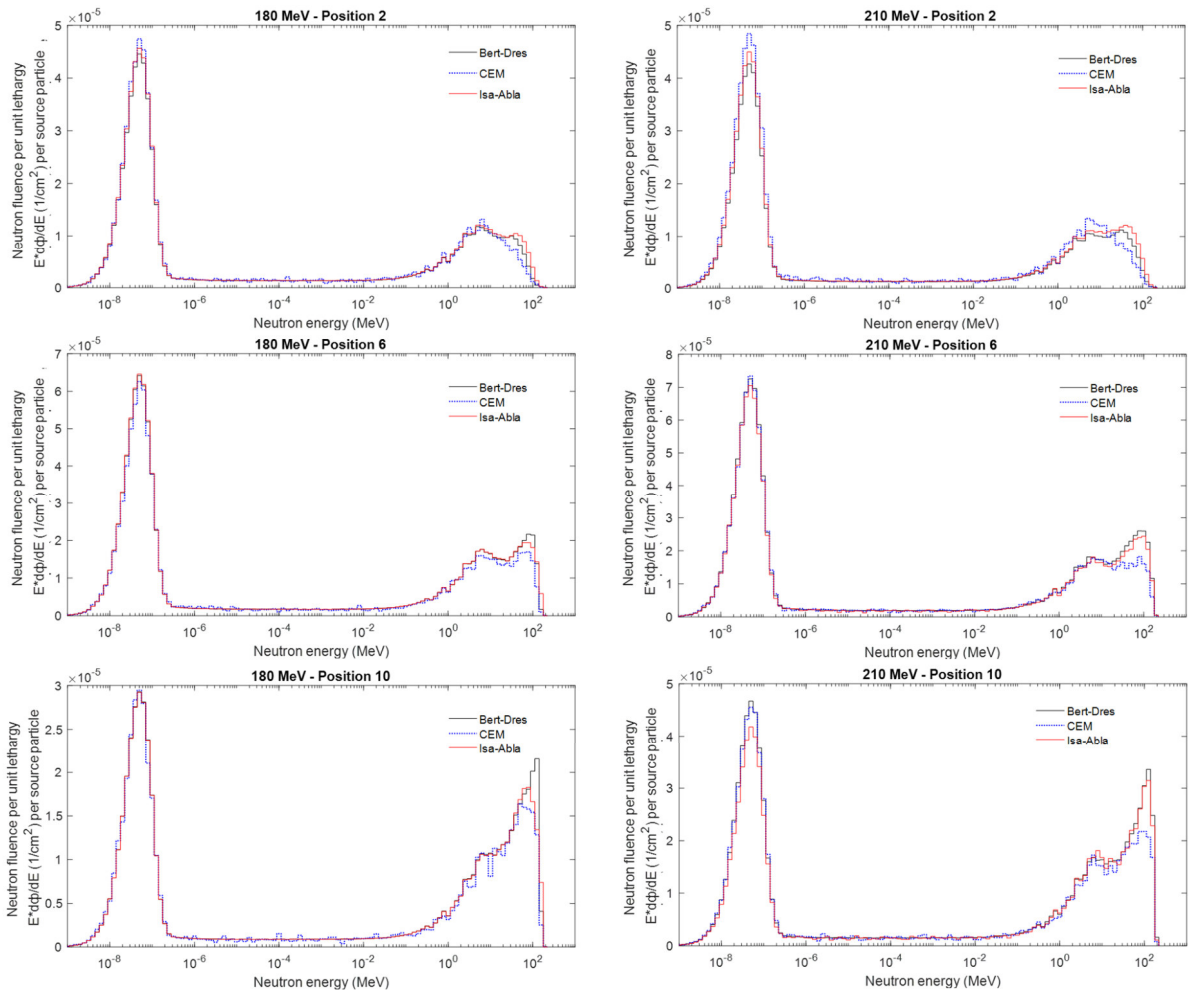

**Figure 4.** Influence of MCNPX nuclear models on neutron fluence spectra simulated at lateral positions 2 (up), 6 (middle) and 10 (down) considering proton beam energies of 180 MeV (left) and 210 MeV (right).

**Influence of MC simulated neutron fluence spectra on the calculation of stray neutron dose equivalent**

Calculations of the neutron dose equivalent (mSv/source particle) as obtained by averaging results from the different codes are shown in table 7 together with variations in thel neutron dose equivalent computed by equation (1) and using neutron spectra of the different MC codes at different distal (5 and 9) and lateral (2, 6 and 9) positions for proton beams of 110 MeV, 150 MeV, 180 MeV and 210 MeV. In distal positions, variations are generally larger compared to lateral positions and reach up to 45% at position 9 for 110 MeV protons. Obviously due to the different energy spectra and shape of the energy dependence of kerma and quality factors, the variations in the total fluence (see table 4 and 5) have a different impact on the equivalent doses (table 7). For 110 MeV protons we see that the variation of the total neutron dose equivalent is larger than the variation of the total neutron fluence. For 210 MeV protons this finding is not always observed due to only minor differences in high energy region of neutron spectra (compare figures 2 and 3). **Table 7.** Average values of neutron dose equivalent from different MC codes (top) and variation of the neutron dose equivalent calculated from spectra simulated with different codes (bottom) in distal positions 5 and 9 and lateral positions 2, 6 and 10 for 110 MeV, 150 MeV, 180 MeV and 210 MeV proton beams.

| Average neutron dose equivalent [mSv per source particle] | | | | |
|---|---|---|---|---|
| Position | 110 MeV | 150 MeV | 180 MeV | 210 MeV |
| 5 | 4.31E-11 | IF | IF | IF |
| 9 | 1.14E-11 | 4.06E-11 | 9.76E-11 | IF |
| 2 | 9.74E-12 | 1.40E-11 | 1.62E-11 | 1.75E-11 |
| 6 | 1.13E-11 | 2.35E-11 | 3.15E-11 | 3.73E-11 |
| 10 | 5.95E-12 | 1.68E-11 | 2.78E-11 | 4.11E-11 |
| Variation (%) of neutron dose equivalent between codes | | | | |
| Position | 110 MeV | 150 MeV | 180 MeV | 210 MeV |
| 5 | 34 | IF | IF | IF |
| 9 | 45 | 32 | 25 | IF |
| 2 | 19 | 24 | 25 | 21 |
| 6 | 12 | 5 | 13 | 15 |
| 10 | 30 | 16 | 7 | 5 |

IF: in field point

In lateral positions, variations of neutron dose equivalent in position 2 were on average 22% for the different proton energies while the same variation in position 6 was generally lower on average around 11% (ranging between 5% and 15%). The largest variation for lateral positions was observed in position 10 for 110 MeV, which reached up to 30%. This can be explained by rather big differences in the high energy region of the neutron spectra (see Figure 2) which is in line with the fact that position 10 has a small angle towards the 110 MeV Bragg peak (35 degrees from the field axis) and so is in a more forward direction than the other lateral positions. Using different models in MCNPX, the maximum observed relative variation for the neutron dose equivalent was of 5% (position 2 for 210 MeV).

**Influence of MC simulated neutron fluence spectra on the calibration factor of CR39 detectors**

The impact of different MC codes and models on the calibration factor of CR39 detectors are shown in table 8 for positions 6 and 10 and for different proton energies. Variation between models was very small and, as expected, for 110 MeV and 150 MeV proton energies it remained within 1%. For higher proton energies, the variation of neutron dose equivalent reached up to 8% for 210 MeV in position 6. Variation between codes was on average 10% and 11% for position 6 and 10, respectively.

**Table 8.** CR39 calibration factors and variation between MCNPX models and MC at lateral positions 6 and 10 for 110 MeV, 150 MeV, 180 MeV and 210 MeV proton beams.

|  | Proton energy | Bert-Dres | CEM | Isa-Abla | Variation models (%) | GEANT4 | FLUKA | Variation codes (%) |
|---|---|---|---|---|---|---|---|---|
| Position 6 | 110 MeV | 3.22E-03 | 3.22E-03 | 3.22E-03 | 0% | 3.60E-03 | 3.91E-03 | 10% |
| | 150 MeV | 3.29E-03 | 3.29E-03 | 3.29E-03 | 0% | 4.00E-03 | 4.21E-03 | 13% |
| | 180 MeV | 3.72E-03 | 3.41E-03 | 3.70E-03 | 5% | 4.05E-03 | 4.35E-03 | 8% |
| | 210 MeV | 3.84E-03 | 3.31E-03 | 3.77E-03 | 8% | 4.05E-03 | 4.58E-03 | 9% |
| Position 10 | 110 MeV | 3.83E-03 | 3.88E-03 | 3.83E-03 | 1% | 4.31E-03 | 4.65E-03 | 10% |
| | 150 MeV | 4.19E-03 | 4.19E-03 | 4.19E-03 | 0% | 4.83E-03 | 5.33E-03 | 12% |
| | 180 MeV | 4.36E-03 | 4.10E-03 | 4.28E-03 | 3% | 5.20E-03 | 5.44E-03 | 11% |
| | 210 MeV | 4.29E-03 | 3.99E-03 | 4.22E-03 | 4% | 4.80E-03 | 5.52E-03 | 13% |

**Discussion**

This study focused on comparing three widely used MC codes, FLUKA, MCNPX and GEANT4 in the prediction of secondary neutrons spectra for the assessment of the neutron dose equivalent as well as for calibration of detectors, such as CR39.

Firstly, the largest differences in calculating the neutron spectra were observed between different codes and most pronounced in the forward beam direction. Variation between the codes reached up to around 50%, with a maximum disagreement up to 200%, for the high energy neutrons in the forward positions 5 and 9 for 110 MeV. Most likely, this was related to the more prominent high energy component in the forward direction and the lack of cross section data in FLUKA and GEANT4 for neutrons above 20 MeV. MCNPX uses nuclear cross section data up to 150 MeV, and the present work has shown that both FLUKA and GEANT4 tend to calculate higher fluences for the high energy neutrons in forward positions up to 200% for FLUKA (110 MeV position 9) and up to 72% for GEANT4 (110 MeV position 5). For both FLUKA and GEANT4 this higher fluence was most pronounced in distal positions (behind the Bragg peak) and to a lesser extent at lateral positions 6 and 10. This can be due to the more forwarded direction of these positions relative to the Bragg peak for 110 MeV protons with, respectively, a 60 and 35 degrees angle from the isocenter. Indeed looking into the neutron spectra for 110 MeV proton beam it is clear that position 10 involves an important contribution from high energy neutrons while this contribution is much smaller at position 2.

In lateral positions, variations between the codes for 210 MeV protons are smaller and less pronounced for the high energy neutrons, reaching up to 20% for the total fluence in position 10. What is noticeable in these lateral positions for 210 MeV protons is that both FLUKA and GEANT4 calculated lower neutron fluence compared to MCNPX.

In general, the performance of the different codes seemed to be related to the relative position and, more specifically, the angle towards the isocenter. In forward directions a higher fluence of FLUKA and GEANT4 is observed, while in lateral positions (i.e. lateral position) a lower fluence is obtained (Table 3) compared to MCNPX. This angular dependence is likely due to the contribution/proportion of high energy neutrons prominent at forward directions. Overall the thermal neutron fluence is always most pronounced for GEANT4. Our result agree with recently published data from *Englbrecht, et al.* where FLUKA and GEANT4 simulations were compared in different positions inside the room for different mono-energetic protons (75, 118, 140 and 200 MeV) hitting a PMMA phantom [20]. In forward direction they report larger high energy neutron fluences for FLUKA while in lateral positions FLUKA shows lower high energy neutron fluences when compared to GEANT4. Also here GEANT4 thermal neutron fluence was always higher compared to FLUKA which is also in line with our data. Unlike this recent paper we did not model any room component in our study and as a result our data lack evaporation neutrons as the probability of evaporation neutrons is low for light nuclei and rise with rising mass number. We do believe the impact of room components to be less significant for the simulations we have performed here, as the results were compared inside the water phantom and not in the room, as was the case for *Englbrecht, et al.* study [20].

Similarly, the neutron dose equivalent proved to involve larger variations at distal positions compared to lateral positions. Overall, increasing the proton energy decreased slightly the variation between the codes which is clearly observed in position 9 and 10, respectively, going from 45% for 110 MeV to 25% for 180 MeV and from 30% for 110 MeV to 5% for 210 MeV. Nevertheless, comparison of positions for different energies is challenging as the relative position changes with respect to the Bragg peak and so does the spectrum and angular distribution. We did compare the variation in position 2 for 110 MeV with position 10 for 210 MeV, as these are almost equivalent positions lateral to the Bragg peak, which showed comparable variations. In fact, a limitation of the study is that for the high energy proton beam (210 MeV) we do not have distal positions as the beam ranged up till the end of the water phantom. Nevertheless, the phantom dimensions were based on those used in previous measurements performed in EURADOS WG9 measurement campaigns [37, 40]. Moreover, in realistic clinical conditions an energy of 210 MeV with a range in water exceeding 28 cm is rarely applied. In addition, due to geometrical reasons, a proton beam is usually not directed along the patient body so that the maximal neutron exposure in forward directions is limited. Therefore, most of out-of-field positions will be lateral to the beam direction as studied in our work. Nevertheless, in the future we would like to involve modelling of clinically relevant beams (Spread Out Bragg Peaks) and evaluate the influence of different MC codes and nuclear models on the neutron fluence spectra and dose quantities from different codes more deeply. Together with experimental benchmarking of the codes this would allow a refined characterization of in vivo neutron field and translation to radiation protection concerns and solutions, involving recommendations for optimal choice of beam parameters.

Besides the impact of the selected MC codes, the influence of the selected neutron model was evaluated in MCNPX. We observed only an influence for the 180 and 210 MeV proton beams and for modeling the high energy neutrons which was up to 23% for 210 MeV in position 2. Nevertheless, the neutron dose equivalent calculation was not changed more than 5% which was considered within the statistical uncertainty for the lateral positions. Interestingly the CEM model showed always lower results than the default model used in MCNPX (Bertini-Dresner) and Isabel-Abla while both Bertini-Dresner and Isabel-Abla show relatively good agreement with each other. Unfortunately, these models could not be validated against measurements, which was beyond the objective of the study. However, EURADOS Working Groups plan to organize future validation experiments which could benchmark models and test their performance for this specific application in proton therapy. What is noticeable though is that MCNPX by default uses Bertini-Dresner, while the latest versions of MCNP6.2 uses by default the CEM03 Cascade-Exciton model. Moreover, the pre-equilibrium model used by CEM03, the so-called "exciton" model, is more extensively developed so it could be considered as a more reliable model. We did not compare the CEM03 model to FLUKA and GEANT4 data and mostly focused to compare the default code setting, but clearly for high proton energies and at lateral positions, we noted overall lower fluences for both FLUKA and GEANT4 than with the default MCNPX model Bertini-Dresner.

Interestingly, for MCNPX we had different participating users which all performed simulations with their own input files and running the different combinations of nuclear models. This allowed us to make sure there is no bias in the way one person treats the problem (definition of geometries, source etc…), as results showed agreement within statistical uncertainties. Unfortunately for both FLUKA and GEANT4 codes were only used by 1 participant which can be considered as a weak point of the current study. Moreover, for FLUKA and GEANT4 we were not able to perform extra calculations for varying physics settings and models due to the very time consuming and demanding programming skills of these codes. Nevertheless, the choice of the binary intra-nuclear cascade (BIC) model within GEANT4 was based on the fact that there is a better agreement between BIC and experimental data compared to the Bertini model. A good agreement has been shown in the paper of Englbrecht to experimental data and also has been described in a recent report from the Geant4 collaboration [20, 57]. As for the FLUKA code, cross-checking with the newest version of the code (FLUKA 2021.2) could be beneficial, especially for various options available through the PHYSICS card activation which opens the perspective for further studies. In our work, however, the beam model used within the FLUKA code was initially benchmarked with the beam model used in GEANT4 code in previous EURADOS WG9 campaigns [37], verified by the experimental data and therefore agreed to be used in this study.

Finally, when using the spectra for the assessment of the calibration factor of CR39 detectors, the secondary neutron spectra calculated by different MC codes and using different high energy nuclear models was within 10% for the different codes in lateral positions 6 and 10 which reached up to a maximum of 13% for 150 MeV in position 6 and 210 MeV in position 10. The impact of different models reached up to a maximum of 8% for 210 MeV in position 10.

This study describes the uncertainty associated with the MC simulated fluence spectra to assess the calibration factor and can be expected to be around 10% for lateral positions. Compared to the uncertainty associated to the fluence to track density conversion from response factors of 25%, the uncertainty is lower. In previous studies however the combined uncertainty, including this uncertainty with those from track density and MC simulated fluence spectra, was estimated to be around 30%. Unfortunately we were not able to make a direct comparison between simulated data (mono-energetic proton beam) and experimental data (Spread out Bragg peak) from the previous measurement campaign [37], but this is definitely interesting and this work will be continued within EURADOS WG9.

**Conclusion**

This study demonstrated a significant dependence of calculated the neutron fluence spectra on the codes used which has an important implication on both the calculated neutron dose equivalent and calibration of CR39 detectors. Use of different nuclear models in MCNPX showed less prominent variations which were only visible for the high energy proton beams and modeling of high energy neutrons, which results in a minor effect on the calculated neutron dose equivalent and calibration of CR39.

**Acknowledgements**

This work was carried out within EURADOS WG9 - Radiation Dosimetry in Radiohterapy. We would like to thank our colleges for discussion about nuclear models and about existing results of neutron dose measurements inside phantoms.

This research was supported in part by PL-Grid Infrastructure for FLUKA calculations.

**References**

[1] Hall EJ. Intensity-modulated radiation therapy, protons, and the risk of second cancers. *Int J Radiat Oncol Biol Phys 2006*; 65: 1-7. DOI: 10.1016/j.ijrobp.2006.01.027
[2] Gottschalk B. Neutron dose in scattered and scanned proton beams: in regard to Eric J. Hall (Int J Radiat Oncol Biol Phys 2006;65:1-7). *Int J Radiat Oncol Biol Phys 2006*; 66: 1594; author reply 5. DOI: 10.1016/j.ijrobp.2006.08.014
[3] Pelliccioni M. Overview of fluence-to-effective dose and fluence-to ambient dose equivalent conversion coefficients for high energy radiation calculated using the FLUKA code. *Radiat Prot Dosim 2000*; 88: 279-97. DOI: DOI 10.1093/oxfordjournals.rpd.a033046
[4] Ferrari A, Ranft J and Sala PR. The FLUKA radiation transport code and its use for space problems. *Phys Medica 2001*; 17: 72-80.
[5] Bohlen TT, Cerutti F, Chin MPW, Fosso A, Ferrari A, Ortega PG, et al. The FLUKA Code: Developments and Challenges for High Energy and Medical Applications. *Nucl Data Sheets 2014*; 120: 211-4. DOI: 10.1016/j.nds.2014.07.049
[6] Waters LS, McKinney GW, Durkee JW, Fensin ML, Hendricks JS, James MR, et al. The MCNPX Monte Carlo radiation transport code. *Aip Conf Proc 2007*; 896: 81-+.
[7] Agostinelli S, Allison J, Amako K, Apostolakis J, Araujo H, Arce P, et al. GEANT4-a simulation toolkit. *Nucl Instrum Meth A 2003*; 506: 250-303. DOI: 10.1016/S0168-9002(03)01368-8
[8] Sato T, Niita K, Matsuda N, Hashimoto S, Iwamoto Y, Furuta T, et al. Overview of particle and heavy ion transport code system PHITS. *Ann Nucl Energy 2015*; 82: 110-5. DOI: 10.1016/j.anucene.2014.08.023
[9] Jiang HY, Wang B, Xu XG, Suit HD and Paganetti H. Simulation of organ-specific patient effective dose due to secondary neutrons in proton radiation treatment. *Phys Med Biol 2005*; 50: 4337-53. DOI: 10.1088/0031-9155/50/18/007
[10] Polf JC and Newhauser WD. Calculations of neutron dose equivalent exposures from range-modulated proton therapy beams. *Phys Med Biol 2005*; 50: 3859-73. DOI: 10.1088/0031-9155/50/16/014
[11] Zheng Y, Newhauser W, Fontenot J, Taddei P and Mohan R. Monte Carlo study of neutron dose equivalent during passive scattering proton therapy. *Phys Med Biol 2007*; 52: 4481-96. DOI: 10.1088/0031-9155/52/15/008
[12] Jarlskog CZ, Lee C, Bolch WE, Xu XG and Paganetti H. Assessment of organ-specific neutron equivalent doses in proton therapy using computational whole-body age-dependent voxel phantoms. *Phys Med Biol 2008*; 53: 693-717. DOI: 10.1088/0031-9155/53/3/012
[13] Taddei PJ, Mirkovic D, Fontenot JD, Giebeler A, Zheng YS, Kornguth D, et al. Stray radiation dose and second cancer risk for a pediatric patient receiving craniospinal irradiation with proton beams. *Phys Med Biol 2009*; 54: 2259-75. DOI: 10.1088/0031-9155/54/8/001

[14] Zheng Y, Fontenot J, Taddei P, Mirkovic D and Newhauser W. Monte Carlo simulations of neutron spectral fluence, radiation weighting factor and ambient dose equivalent for a passively scattered proton therapy unit. *Phys Med Biol 2008*; 53: 187-201. DOI: 10.1088/0031-9155/53/1/013

[15] De Smet V, De Saint-Hubert M, Dinar N, Manessi GP, Aza E, Cassell C, et al. Secondary neutrons inside a proton therapy facility: MCNPX simulations compared to measurements performed with a Bonner Sphere Spectrometer and neutron H*(10) monitors. *Rad Meas 2017*; 99: 25-40. DOI: https://doi.org/10.1016/j.radmeas.2017.03.005

[16] Schneider U, Agosteo S, Pedroni E and Besserer J. Secondary neutron dose during proton therapy using spot scanning. *Int J Radiat Oncol 2002*; 53: 244-51. DOI: Pii S0360-3016(01)02826-7 Doi 10.1016/S0360-3016(01)02826-7

[17] Zheng Y, Newhauser W, Klein E and Low D. Angular Distribution of Neutron Fluence and Its Effect On Shielding for a Passively-Scattered Proton Therapy Unit. *Med Phys 2008*; 35. DOI: 10.1118/1.2962075

[18] Perez-Andujar A, Newhauser WD and DeLuca PM. Contribution to Neutron Fluence and Neutron Absorbed Dose from Double Scattering Proton Therapy System Components. *Nucl Technol 2009*; 168: 728-35. DOI: Doi 10.13182/Nt09-A9297

[19] Hohmann E, Safai S, Bula C, Luscher R, Harm C, Mayer S, et al. INVESTIGATION OF THE NEUTRON STRAY RADIATION FIELD PRODUCED BY IRRADIATING A WATER PHANTOM WITH 200-MeV PROTONS. *Nucl Technol 2011*; 175: 77-80. DOI: Doi 10.13182/Nt11-A12273

[20] Englbrecht FS, Trinkl S, Mares V, Ruhm W, Wielunski M, Wilkens JJ, et al. A comprehensive Monte Carlo study of out-of-field secondary neutron spectra in a scanned-beam proton therapy gantry room. *Z Med Phys 2021*. DOI: 10.1016/j.zemedi.2021.01.001

[21] Tayama R, Fujita Y, Tadokoro M, Fujimaki H, Sakae T and Terunuma T. Measurement of neutron dose distribution for a passive scattering nozzle at the Proton Medical Research Center (PMRC). *Nucl Instrum Meth A 2006*; 564: 532-6. DOI: 10.1016/j.nima.2006.04.028

[22] Farah J, Martinetti F, Sayah R, Lacoste V, Donadille L, Trompier F, et al. Monte Carlo modeling of proton therapy installations: a global experimental method to validate secondary neutron dose calculations. *Phys Med Biol 2014*; 59: 2747-65. DOI: 10.1088/0031-9155/59/11/2747

[23] Farah J, Sayah R, Martinetti F, Donadille L, Lacoste V, Herault J, et al. Secondary Neutron Doses in Proton Therapy Treatments of Ocular Melanoma and Craniopharyngioma. *Radiat Prot Dosim 2014*; 161: 363-7. DOI: 10.1093/rpd/nct283

[24] Stolarczyk L, Cywicka-Jakiel T, Horwacik T, Olko P, Swakon J and Waligorski MPR. Evaluation of risk of secondary cancer occurrence after proton radiotherapy of ocular tumours. *Rad Meas 2011*; 46: 1944-7. DOI: https://doi.org/10.1016/j.radmeas.2011.05.046

[25] Mosteller RD, Frankle SC and Young PG, *Data Testing of ENDF/B-VI with MCNP: Critical Experiments, Thermal-Reactor Lattices, and Time-of-Flight Measurements*, in Advances in Nuclear Science and Technology, J. Lewins and M. Becker, Editors. 1997, Springer US: Boston, MA. p. 131-95.

[26] Chadwick MB, Obložinský P, Herman M, Greene NM, McKnight RD, Smith DL, et al. ENDF/B-VII.0: Next Generation Evaluated Nuclear Data Library for Nuclear Science and Technology. *Nucl Data Sheets 2006*; 107: 2931-3060. DOI: https://doi.org/10.1016/j.nds.2006.11.001

[27] Little RC, Frankle SC, Hughes IHG and Prael RE. Utilization of new 150-MeV neutron and proton evaluations in MCNP. 1997. United States.

[28] Ferrari A, La Torre FP, Manessi GP, Pozzi F and Silari M. Spallation cross sections for nat Fe and nat Cu targets for 120 GeV/c protons and pions. *Physical Review C 2014*; 0346128980. DOI: 10.1103/PhysRevC.89.034612

[29] Wagner V, Suchopár M, Vrzalová J, Chudoba P, Svoboda O, Tichý P, et al. How to Use Benchmark and Cross-section Studies to Improve Data Libraries and Models. *Journal of Physics: Conference Series 2016*; 724: 012052. DOI: 10.1088/1742-6596/724/1/012052

[30] Pioch C, Mares V and Rühm W. Influence of Bonner sphere response functions above 20 MeV on unfolded neutron spectra and doses. *Radiat Meas 2010*; 45: 1263-7. DOI: https://doi.org/10.1016/j.radmeas.2010.05.007

[31] Rühm W, Mares V, Pioch C, Agosteo S, Endo A, Ferrarini M, et al. Comparison of Bonner sphere responses calculated by different Monte Carlo codes at energies between 1 MeV and 1 GeV – Potential impact on neutron dosimetry at energies higher than 20 MeV. *Radiat Meas 2014*; 67: 24-34. DOI: https://doi.org/10.1016/j.radmeas.2014.05.006

[32] Romero-Expósito M, Domingo C, Sánchez-Doblado F, Ortega-Gelabert O and Gallego S. Experimental evaluation of neutron dose in radiotherapy patients: Which dose? *Med Phys 2016*; 43: 360. DOI: 10.1118/1.4938578

[33] Hälg R, Besserer J, Boschung M, Mayer S, Clasie B, Kry S, et al. Field calibration of PADC track etch detectors for local neutron dosimetry in man using different radiation qualities. *Nuclear Instruments and Methods in Physics Research Section A: Accelerators, Spectrometers, Detectors and Associated Equipment 2012*; 694: 205–10. DOI: 10.1016/j.nima.2012.08.021

[34] Mares V, Romero-Expósito M, Farah J, Trinkl S, Domingo C, Dommert M, et al. A comprehensive spectrometry study of a stray neutron radiation field in scanning proton therapy. *Phys Med Biol 2016*; 61: 4127-40. DOI: 10.1088/0031-9155/61/11/4127

[35] Trinkl S, Mares V, Englbrecht FS, Wilkens JJ, Wielunski M, Parodi K, et al. Systematic out-of-field secondary neutron spectrometry and dosimetry in pencil beam scanning proton therapy. *Med Phys 2017*; 44: 1912-20. DOI: 10.1002/mp.12206

[36] Farah J, Mares V, Romero-Exposito M, Trinkl S, Domingo C, Dufek V, et al. Measurement of stray radiation within a scanning proton therapy facility: EURADOS WG9 intercomparison exercise of active dosimetry systems. *Med Phys 2015*; 42: 2572-84. DOI: 10.1118/1.4916667

[37] Stolarczyk L, Trinkl S, Romero-Expósito M, Mojżeszek N, Ambrozova I, Domingo C, et al. Dose distribution of secondary radiation in a water phantom for a proton pencil beam—EURADOS WG9 intercomparison exercise. *Physics in Medicine & Biology 2018*; 63: 085017. DOI: 10.1088/1361-6560/aab469

[38] Wochnik A, Stolarczyk L, Ambrožová I, Davídková M, De Saint-Hubert M, Domański S, et al. Out-of-field doses for scanning proton radiotherapy of shallowly located paediatric tumours-a comparison of range shifter and 3D printed compensator. *Phys Med Biol 2021*; 66: 035012. DOI: 10.1088/1361-6560/abcb1f

[39] Knežević Ž, Ambrozova I, Domingo C, De Saint-Hubert M, Majer M, Martínez-Rovira I, et al. COMPARISON OF RESPONSE OF PASSIVE DOSIMETRY SYSTEMS IN SCANNING PROTON RADIOTHERAPY-A STUDY USING PAEDIATRIC ANTHROPOMORPHIC PHANTOMS. *Radiat Prot Dosim 2018*; 180: 256-60. DOI: 10.1093/rpd/ncx254

[40] Bordy JM, Bessieres I, d'Agostino E, Domingo C, d'Errico F, di Fulvio A, et al. Radiotherapy out-of-field dosimetry: Experimental and computational results for photons in a water tank. *Rad Meas 2013*; 57: 29-34. DOI: https://doi.org/10.1016/j.radmeas.2013.06.010

[41] Folger G, Ivanchenko VN and Wellisch J. The Binary Cascade: Nucleon nuclear reactions. *The European Physical Journal A - Hadrons and Nuclei 2004*; 21: 407-17. DOI: 10.1140/epja/i2003-10219-7

[42] Pelowitz DB, MCNPX USER'S MANUAL: Version 2.7.0, in Los Alamos National Laboratory report LA-CP-11-00438. 2011.

[43] Chadwick MB. Neutron, proton, and photonuclear cross-sections for radiation therapy and radiation protection. *Radiat Environ Bioph 1998*; 37: 235-42. DOI: DOI 10.1007/s004110050124

[44] Chadwick MB, Barschall HH, Caswell RS, DeLuca PM, Hale GM, Jones DTL, et al. A consistent set of neutron kerma coefficients from thermal to 150 MeV for biologically important materials. *Medical Physics 1999*; 26: 974-91. DOI: https://doi.org/10.1118/1.598601

[45] Chadwick MB. Nuclear reactions in proton, neutron, and photon radiotherapy. *Radiochim Acta 2001*; 89: 325-36. DOI: DOI 10.1524/ract.2001.89.4-5.325

[46] Bertini HW. *Nucl. Instr. and Meth 1968*; 66.

[47] Dresner LW. *Oak Ridge Report ORNL-TM-196 1962*.

[48] Yariv Y. *Phys. Rev. 1969*; 188.

[49] Gudima KK, Mashnik SG and Toneev VD. Cascade-exciton model of nuclear reactions. *Nuclear Physics A 1983*; 401: 329-61. DOI: https://doi.org/10.1016/0375-9474(83)90532-8

[50] Kerby LM and Mashnik SG. Total reaction cross sections in CEM and MCNP6 at intermediate energies. *Nuclear Instruments and Methods in Physics Research Section B: Beam Interactions with Materials and Atoms 2015*; 356-357: 135-45. DOI: https://doi.org/10.1016/j.nimb.2015.04.057

[51] Robert C, Dedes G, Battistoni G, Bohlen TT, Buvat I, Cerutti F, et al. Distributions of secondary particles in proton and carbon-ion therapy: a comparison between GATE/Geant4 and FLUKA Monte Carlo codes. *Phys Med Biol 2013*; 58: 2879-99. DOI: 10.1088/0031-9155/58/9/2879

[52] FLUKAWebsite. 2005.

[53] Ferrari A and Sala PR. Nuclear reactions in Monte Carlo codes. *Radiat Prot Dosim 2002*; 99: 29-38. DOI: DOI 10.1093/oxfordjournals.rpd.a006788

[54] Sorge H, Stöcker H and Greiner W. Relativistic quantum molecular dynamics approach to nuclear collisions at ultrarelativistic energies. *Nuclear Physics A 1989*; 498: 567-76. DOI: https://doi.org/10.1016/0375-9474(89)90641-6

[55] Cerutti F, Ballarini F, Battistoni G, Colleoni P, Ferrari A, Förtsch SV, et al. Carbon induced reactions at low incident energies. *Journal of Physics: Conference Series 2006*; 41: 212-8. DOI: 10.1088/1742-6596/41/1/021

[56] Siebert B and Schuhmacher H. Quality factors, ambient and personal dose equivalent for neutrons, based on the new ICRU stopping power data for protons and alpha particles. *Radiat Prot Dosim 1995*; 58: 177-83.

[57] Arce P, Bolst D, Bordage MC, Brown JMC, Cirrone P, Cortés-Giraldo MA, et al. Report on G4-Med, a Geant4 benchmarking system for medical physics applications developed by the Geant4 Medical Simulation Benchmarking Group. *Med Phys 2021*; 48: 19-56. DOI: 10.1002/mp.14226